\documentclass[%
 reprint,
superscriptaddress,
longbibliography,
 amsmath,amssymb,
 aps,
prb,
]{revtex4-1}
\setcitestyle{numbers,square}

\usepackage{braket}
\usepackage{graphicx}
\usepackage{dcolumn}
\usepackage{bm}
\usepackage[normalem]{ulem} 
\usepackage{color}
\usepackage{siunitx}
\usepackage[version=4]{mhchem}

\DeclareSIUnit\elementarycharge{\text{\ensuremath{e}}}
\DeclareSIUnit\angstrom{\text {Å}}
\DeclareSIUnit\uc{\text {unit-cell}}

\usepackage{hyperref}
\hypersetup{
    colorlinks,%
    citecolor=blue,%
    linkcolor=blue,%
    urlcolor=blue
}

\newcommand{\orcid}[1]{\href{https://orcid.org/#1}{\includegraphics[width=8pt]{orcid.pdf}}}

\begin{document}

\title{Electronic and spin transport in Bismuthene with magnetic impurities}

\author{Armando Pezo}
\affiliation{Aix-Marseille Univetsité, CNRS, CINaM, Marseille, France}

\author{Felipe Crasto de Lima} 
\affiliation{Ilum School of Science, CNPEM, Campinas, Brazil}

\author{Adalberto Fazzio}
\email{adalberto.fazzio@ilum.cnpem.br}
\affiliation{Ilum School of Science, CNPEM, Campinas, Brazil}

\date{\today}

\begin{abstract}
Topological insulators have remained as candidates for future electronic devices since their first experimental realization in the past decade. The existence of topologically protected edge states could be exploited to generate a robust platform and develop quantum computers. In this work we explore the role of magnetic impurities in the transport properties of topological insulators, in particular, we study the effect on the edge states conductivity. By means of realistic $\it{ab}$ $\it{initio}$ calculations we simulate the interaction between magnetic adatoms and topological insulators, furthermore, our main goal is to obtain the transport properties for large samples as it would be possible to localize edge states at large scales.

\end{abstract}

\maketitle

\section{Introduction}

Topological materials brought new possibilities in the development of new technologies, particularly in spin-based electronic devices. After the theoretical predictions \cite{kane_mele_qshe_graphene,Bernevig1757} they were experimentally observed by means of transport measurements \cite{Konig}. Subsequent experimental characterization was mainly based on angle-resolved photoemission spectroscopy (ARPES) techniques \cite{ando_review}. However, performing electronic transport experiments faces several difficulties related to the (i) substrate, (ii) temperature effects and (iii) topological energy gap. Surpassing these three difficulties one of the main candidates to realize the quantum spin Hall effect (QSHE) at room-temperature is bismuthene \cite{Reis2017}. Due to its large Spin Orbit Coupling (SOC), bismuthene and Bi-based materials serve as good candidates for spintronic and  valleytronic applications \cite{Ji2016, Lu2017, Zhang2015, Liu2017, Wang2017_1}. 

A great deal of interest in topological insulators arise given the Bulk-Boundary correspondence \cite{essin_bulk_boundary, prodan_2016, mong_edge_state, asboth_2016}, which relates the non-trivial bulk band topology to the existence of metallic surface states through a topological invariant \cite{z2_KaneMele.95.146802}. Within the class of different topological phases, the quantum spin Hall (QSH) effect, arising in a finite 2D system, present a spin-momentum locking for the edge states  \cite{zhao2020determination, PhysRevB.91.245112,Ortiz_2016}. However, impurities and adatoms in the system are shown to modify this edge state character \cite{doi_10.1021_nl5009212}, but retaining the topological protection against backscattering for impurities not mixing the edge state spins. The spin-momentum locking dictate the edge states spins aligned with the direction perpendicular to the nanoribbon plane being up/down depending on the edge state momentum velocity positive/negative. Only impurities with magnetic axis not aligned with such spin direction can introduce spin-mixing terms \cite{PhysRevB.106.245408,Novelli2019}. 

This scattering is consistent with the field theoretical description of the 1-dimensional edge states, which experience a perturbation in terms of the magnetic impurity moment modeled by $H'= J\vec{\sigma} \cdot \mathbf{S}\delta(x-x')$ \cite{PhysRevB.1.4464, yazdani_back, helical_liquid_wu, kondo_2009}, being $J$ the strength of the interaction between the helical edge states and a magnetic moment with spin $\mathbf{S}$. It is worth noticing that this perturbation acts only locally, so it's expected that this interaction is spatially limited to a region surrounding the magnetic adatom. Therefore one could expect that magnetic adatoms will have an impact whenever they participate on measurements of realistic samples. 

In this paper, we study the electronic behaviour of the interaction between topological insulators and adatoms using transport simulations, in particular our focus is related to magnetic adatoms (breaking time reversal symmetry), then allowing back-scattering between helical states at the same edge. We carry out density functional theory (DFT) calculations and perform transport calculations through non-equilibrium Green's functions (NEGFs) using as inputs the converged \textit{ab initio} Hamiltonians. A decimation approach is employed which allows the possibility to construct our scattering region from different small building blocks containing magnetic adatoms located randomly along the material.

\section{Methods}

We use a plane-wave basis set to obtain the optimized structures along with a localized basis for the electronic transport. In both cases we used the Perdew-Burke-Ernzerhof \cite{gga,pbe} exchange-correlation functional. We performed the geometry optimizations with the plane-wave basis as implemented in the Vienna \textit{Ab-initio} Simulation Package (VASP) \cite{vasp1, vasp2}. For the simulations, we employ 350\,eV for the plane-wave expansion cutoff. The ionic potentials were described using the projector augmented-wave (PAW) method \cite{paw} with a force convergence criteria of 0.005\,eV/{\AA}.

The transport calculations were performed using the Hamiltonian matrix obtained directly from SIESTA \cite{siesta_method}, using a single-zeta plus polarization basis. We used a real space mesh cutoff energy of 350\,Ry, sampling the reciprocal space with 10 $\vec{k}$-points along the periodic direction of our nanoribbons. We added 20\,{\AA} of vacuum for both non-periodic directions and the nanoribbon edges were hydrogen-passivated. The SOC is introduced using the on-site approximation \cite{siesta_on-site_soc}. The Hamiltonian ($\hat{H}$) and overlap ($\hat{S}$) matrices are obtained after performing a full self-consistent cycle.

\begin{figure}[!t]
    \centering
    \includegraphics[width=\linewidth]{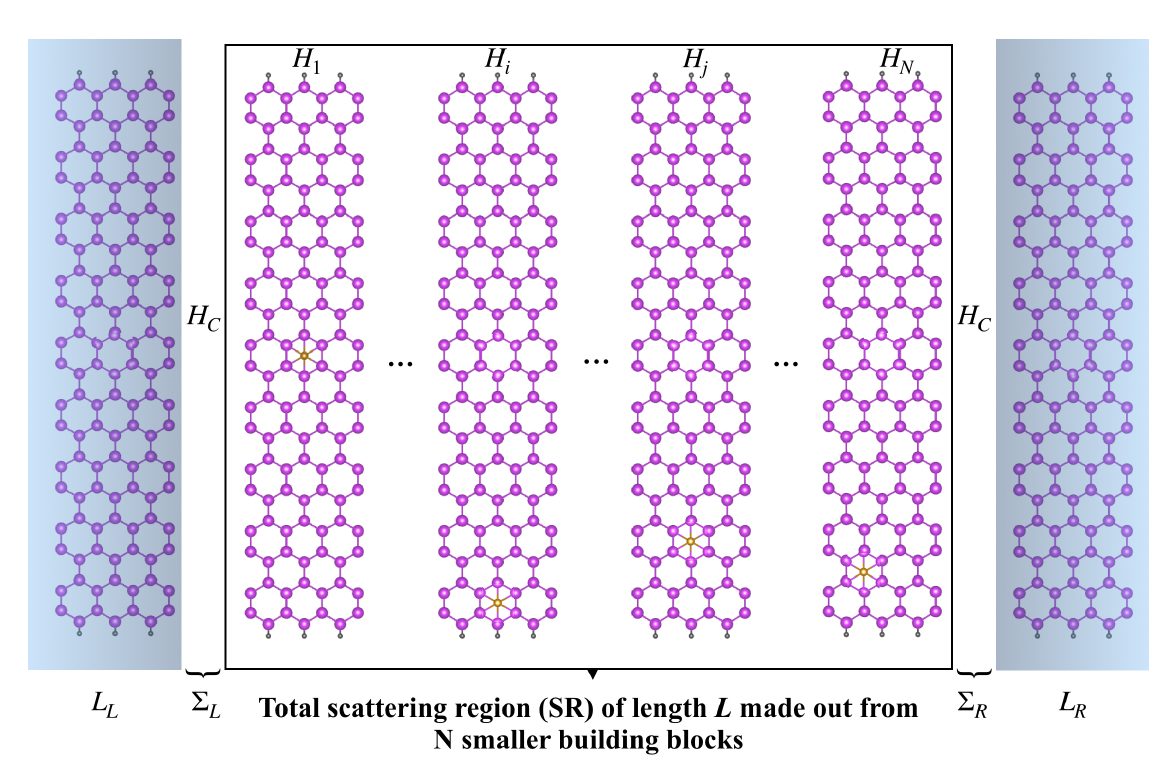}
    \caption{Schematic representation of the two-probe setup studied in this work. The electrodes are located at the left ($L_L$) and right ($L_R$) sides of the device. The scattering region (SR) with length $L$, made out from building blocks ($H_i$) containing adatoms and coupled between each other ($H_C$), is sandwiched by the electrodes which effective interactions are represented by the self energies $\Sigma_L$ and $\Sigma_R$.}
    \label{fig:device}
\end{figure}

The NEGFs electronic transport problem consists in solving the Hamiltonian below~\cite{Caroli_1971}
\begin{equation}
\hat{H} = \begin{pmatrix}
\hat{H}_L & \hat{H}_C & 0 & \hdots & 0 \\
\hat{H}^\dagger_C & \hat{H}_i & \hat{H}_C & 0 &\vdots \\
 0 & 0 & \ddots &\vdots &0 \\
 \vdots & \vdots & \hat{H}^\dagger_C &  \hat{H}_j &\hat{H}_C \\
0 & \hdots & 0 & \hat{H}^\dagger_C & \hat{H}_R \\
\end{pmatrix}\label{eq:big_hamiltonian}
\end{equation}
where $\hat{H}_L$ and $\hat{H}_R$ are the Hamiltonian matrices describing the electrodes while $\hat{H}_C$ are the coupling between the leads and the central region, and $\hat{H}_i$'s are the matrices forming the total scattering region (SR). The SR Green's function \cite{datta-transport,data-negf} is given by
\begin{equation}
    G^{Ret}_M(E)=(\epsilon^+ \hat{S}_M -\hat{H}_M -\Sigma_L^{Ret}(E) -\Sigma_R^{Ret}(E))^{-1},
\end{equation}
where $\Sigma_{L,R}$ are the self-energies for left and right electrodes. In Fig. \ref{fig:device} is depicted a schematic representation of the transport setup.

Our main goal consists in the study of a sample containing a large number of adatoms, to achieve this, we must increase the system's size making exact diagonalization of \eqref{eq:big_hamiltonian} computationally expensive. For this reason, the decimation technique is a valid approach that was successfully applied previously \cite{PhysRevMaterials.5.014204,rossi_transport,transport-carbon-nano,deci_james,Pezo2019}. In such case, the scattering region is divided into building blocks connected via first neighbors' interactions. Each building block is computed through DFT and the scattering region Green's function is obtained after the decimation, then it takes into account the degrees of freedom for the whole system comprised by all the building blocks. The interaction with the leads is introduced via the self-energies written like $\Sigma^{Ret}(E)=(\epsilon^+ \hat{S}-\hat{H})G^{0(Ret)}(E)(\epsilon^+\hat{S}-\hat{H})$ once the surface Green's functions ($G^{0(Ret)}$) are obtained \cite{Sancho_1985,tb-green}. The transmission can be calculated for both spin-conserved and spin-flip parts,
 \begin{equation}
     T^{\sigma \sigma'}=Tr[\Gamma_L^{\sigma \sigma'}(G_M^{\sigma \sigma'})^{\dagger}\Gamma^{\sigma \sigma'}_RG_M^{\sigma \sigma'}],
\end{equation}
where the coupling matrices are $\Gamma^{\sigma,\sigma'}_{L/R}={i}[\Sigma^{(Ret) \sigma,\sigma'}_{L/R}-\Sigma^{(Ret)\dagger \sigma,\sigma'}_{L/R}]$,such that the total transmission $T(E)$ is written like
\begin{equation}
    T(E)=\sum_{\sigma,\sigma'} T^{\sigma \sigma'},
\end{equation}
from which we arrive to the conductance by \cite{fisher_lee_cond_transm, landauer}
\begin{equation}
    G={\frac{e^2}{h}}T(E_F),
\end{equation}
in units of $G_0=e^2/h$.

In order to gain more information for the Spin transport we define the following polarization \cite{spin_coherence_pareek,wudmir_spin_coherence}
\begin{equation}
P(E)=\frac{T^{\sigma,\sigma}-T^{\sigma,\sigma'}} {T^{\sigma,\sigma}+T^{\sigma,\sigma'}}, \label{eq:pol} 
\end{equation}
where we evaluate the normalized difference between the spin-conserved and the the spin-flip transmission, such that the polarization can take values within the window $[-1,1]$, being $-1$ the value corresponding to the case where we have a maximum spin-flip transmission whereas $+1$ means that we had a full spin-conserved transmission.

\section{Results and Discussion}

\subsection{Bismuthene and single adatom}

\begin{figure}
\includegraphics[width=\linewidth]{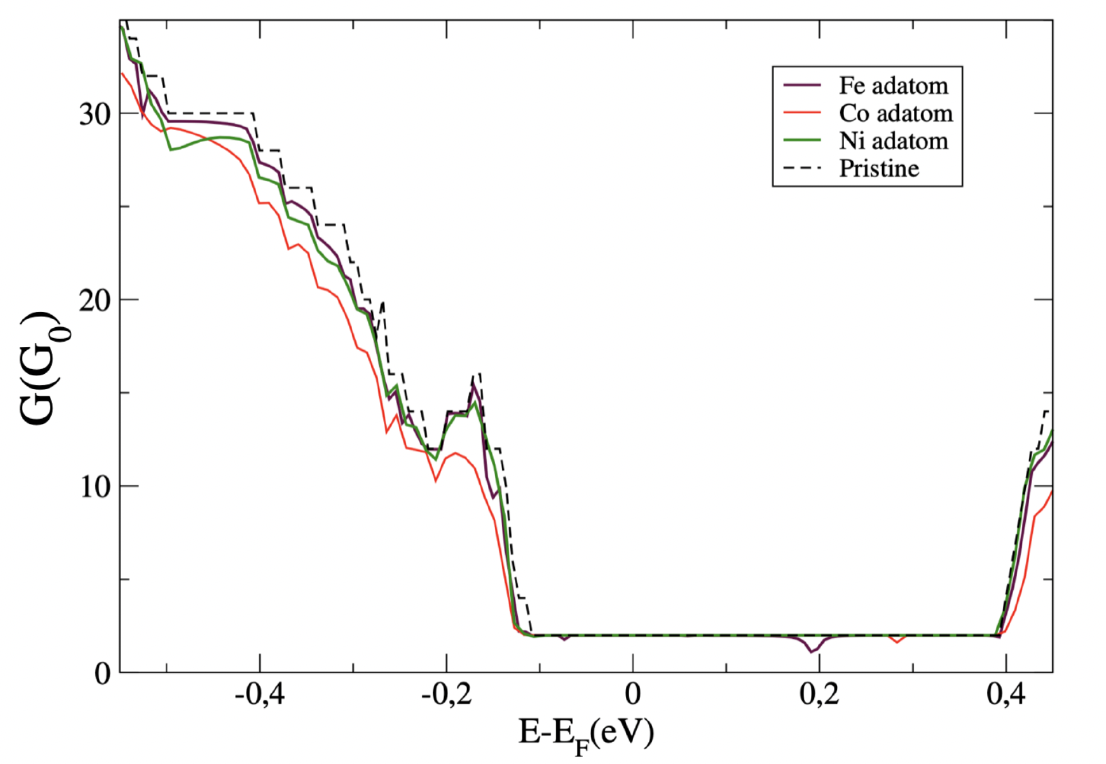}
\caption{Conductance for three different single adatoms:Fe, Co and Ni, deposited along one edge of the nanoribbon. The drop in conductance reflects the fact that we have different magnetic moments generated for each of the three cases. We must emphasize that our simulations give no resultant magnetic moment for the Ni adatom.}\label{fig:conductance_one_adatom}
\end{figure}

\begin{figure}
\includegraphics[width=\linewidth]{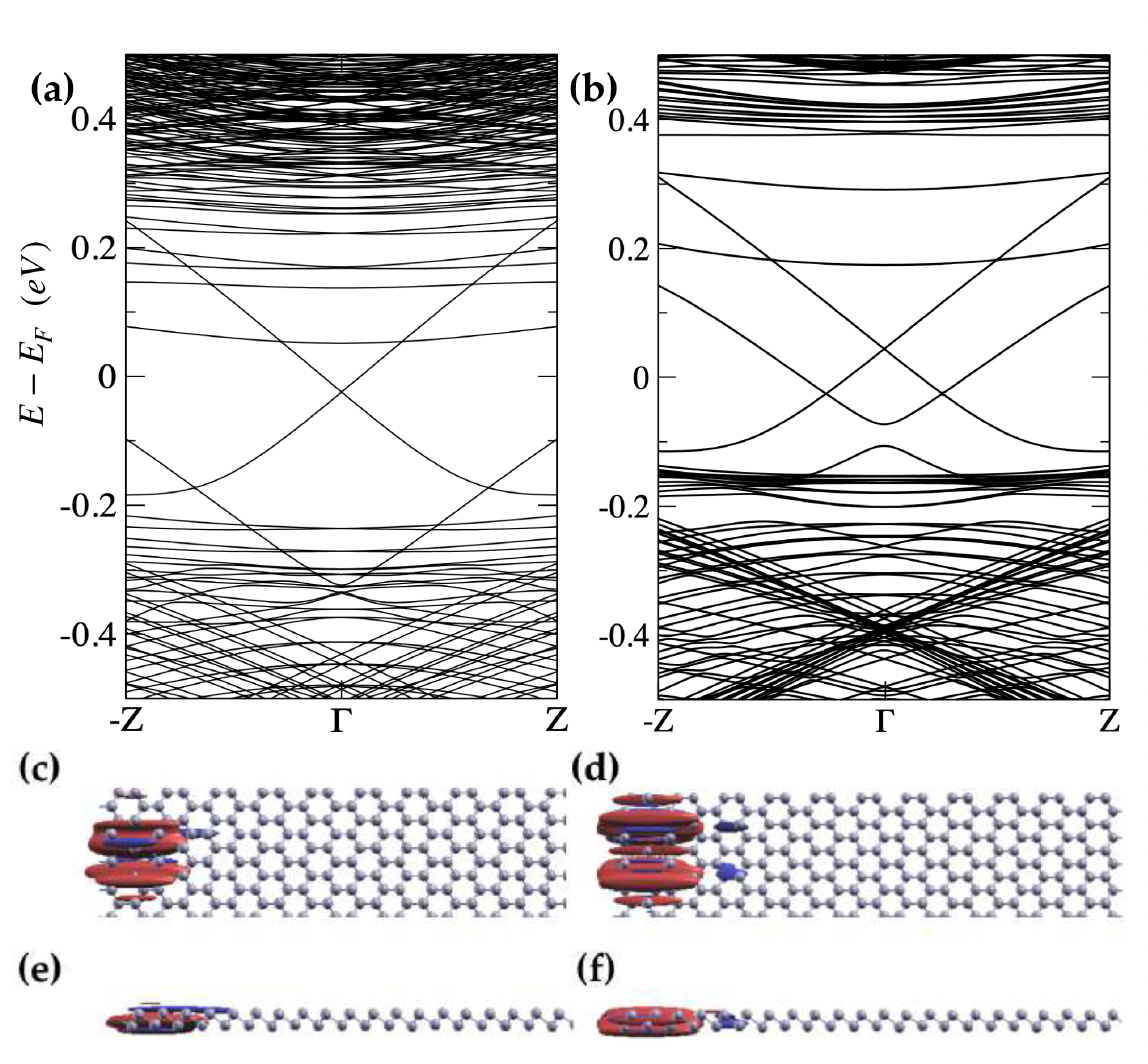}
\caption{Bands (a), top (c) and side-views (e) of the LDOS for bismuthene nanoribbons with a Co adatom at one edge. Analogously, the same plots for the case of Fe adatom are depicted in (b),(d) and (f) respectively.}\label{fig:bands_fe_co_ldos}
\end{figure}

In Fig. \ref{fig:conductance_one_adatom} we show the conductance obtained for three different adatoms on the ribbon edge. Only Nickel adatom has no effect on the condutance within the in the bulk energy gap, given its non-magnetic ground state. This means that Nickel does not break time-reversal and backscattering protection is preserved. Magnetic Co and Fe adatoms present energy values with non-quantized electronic transport, with the larger drop of the conductance in the case of Iron giving its larger magnetic moment. The strength of magnetic moment and adsorption energy are presented in Table \ref{table:mag_one_moment_bis}. The adsorption energies ($E_{ads}$) are defined as
\begin{equation*}
    E_{ads}=E_{ribbon+ad}-E_{ribbon}-E_{ad},
\end{equation*}
where $E_{ribbon+ad}$ is the energy of the ribbon containing one adatom, $E_{ribbon}$ is the pristine ribbon energy and $E_{ad}$ is the adatom isolated energy. Such values are consistent with previous works \cite{PhysRevB.96.245424}, and their negative $E_{ads}$, indicate exothermic process, being possible contaminants in experimental systems. In Fig. \ref{fig:bands_fe_co_ldos} (a) and (b) are depicted the bands for bismuthene ribbons containing Co and Fe adatoms respectively. Here we can see (i) the presence of mid-gap states come from the magnetic adatoms (ii) and energy shift between left/right edge states given the adatom charge transfer, and (iii) a larger gap opening in the case of the Fe atom. The LDOS for the mid-gap states are shown in Fig.~\ref{fig:bands_fe_co_ldos} panels (c)-(f) indicating the localized nature on the adatom.

\begin{table}
\caption{\label{table:mag_one_moment_bis} Table showing the magnetic moment ($m$) and adsorption energy ($E_a$) adatom of Ni, Co and Fe. The magnetic moment axis was found to be aligned to the transport direction.}
\begin{ruledtabular}
 \begin{tabular}{ccc} 
 Adatom & $m$ ($\mu_B$) & $E_a$ (eV)\\ 
 \hline
 Ni  & 0.0  &  -2.10 \\ 
 Co  & 2.11 &  -2.45 \\
 Fe  & 2.87 &  -2.81 \\
 \end{tabular}
 \end{ruledtabular}
\end{table}

From Fig. \ref{fig:bands_fe_co_ldos} we note how the edge states dispersion relation changes after introducing the magnetic adatom, such an effect can be explained by considering the following effective Hamiltonian
\begin{equation}
\begin{split}
    \hat{H}_{\rm eff} = \hbar v_F k \sigma_z \otimes \tau_z + \\ (\vec{m}\cdot\vec{\sigma} \delta(x-x')) \otimes \frac{(\tau_z+\tau_0)}{2} + \\ V\sigma_0 \otimes \tau_z,
\end{split}
\end{equation}
being $\sigma$ and $\tau$ the Pauli matrices corresponding to spin and edge spaces, $v_F$ the Fermi velocity, $k$ the momentum, $\vec{m}$ the adatom magnetic moment, $x'$ the adatom position, $V$ the potential difference between left/right edge states. Such a model, in the case of a purely out-of-plane magnetization, leads to an energetic shift due to the $m_z$ component of the magnetization, whereas the gap opening is an outcome of the in-plane magnetization component which will be proportional to either $m_x$ or $m_y$ depending on the momentum direction of the edge state.

\begin{figure}[h]
    \centering
    \includegraphics[width=\columnwidth]{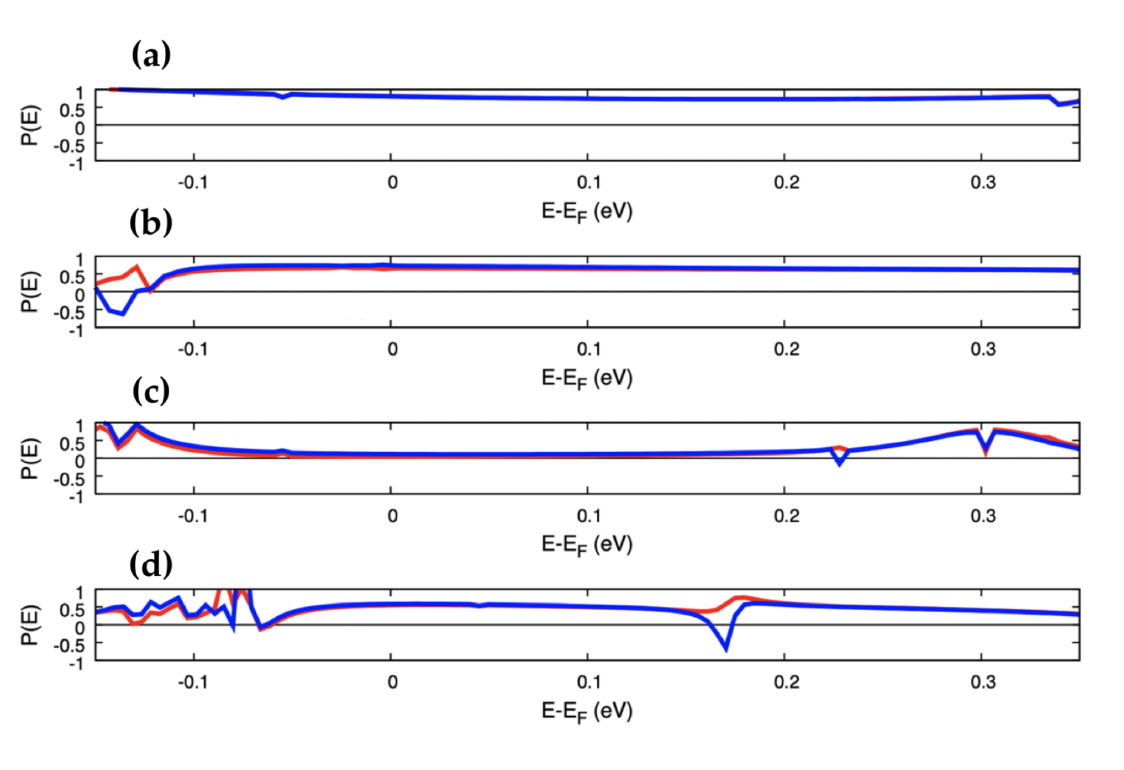}
    \caption{Polarization from Eq.~\ref{eq:pol} for the (a) pristine, (b) Ni, (c) Co, and (d) Fe adatoms on one edge. Positive (Negative) values show a spin-conserving (spin-flip) dominant transport.}
    \label{fig:polarization_one_adatom}
\end{figure}

Giving such effective Hamiltonian interpretation we can have spin-flip effects for (i) adatoms with in-plane magnetization, (ii) inter-edge coupling. In Eq.~\ref{eq:pol} we indicate the Polarization as a character of spin-flip/spin-conserved transmission. In Fig.~\ref{fig:polarization_one_adatom} we present the Polarization for pristine and edge adsorbed adatoms. We can see that for pristine system [Fig.~\ref{fig:polarization_one_adatom}(a)] spin-conserving transport is dominant ($P(E) \sim 1$) within the whole topological energy gap. The residual spin-flip appearing in $P(E)$ is a signature of the edge-edge scattering give the finite size of the nanorribon width ($\sim 80$\,{\AA}). For the non-magnetic Ni adatom, similar behavior is observed [Fig.~\ref{fig:polarization_one_adatom}(b)], where the presence of the Ni impurity mediate a greater edge-edge interaction, leading to a small spin-flip process (however greater than the pristine). For the magnetic impurities Co and Fe [Fig.~\ref{fig:polarization_one_adatom}(c) and (d)] the scenario is drastically changed. Despite the charge transport being quantized within most of the topological energy gap [Fig.~\ref{fig:conductance_one_adatom}] significant spin-flip mechanisms are present. For instance, on Co system the spin-flip and spin-conserving mechanism present similar contributions leading to $P(E) \sim 0$. For Fe adatom in most of the energy range, spin-conserving mechanism dominates over spin-flip ($P(E) \sim 0.5$), however for energy in resonance with the localized Fe impurity energy level [Fig.~\ref{fig:bands_fe_co_ldos} (b)] the spin-flip transport is dominant.

\subsection{Localization effect}

\begin{figure}
    \includegraphics[width=\linewidth]{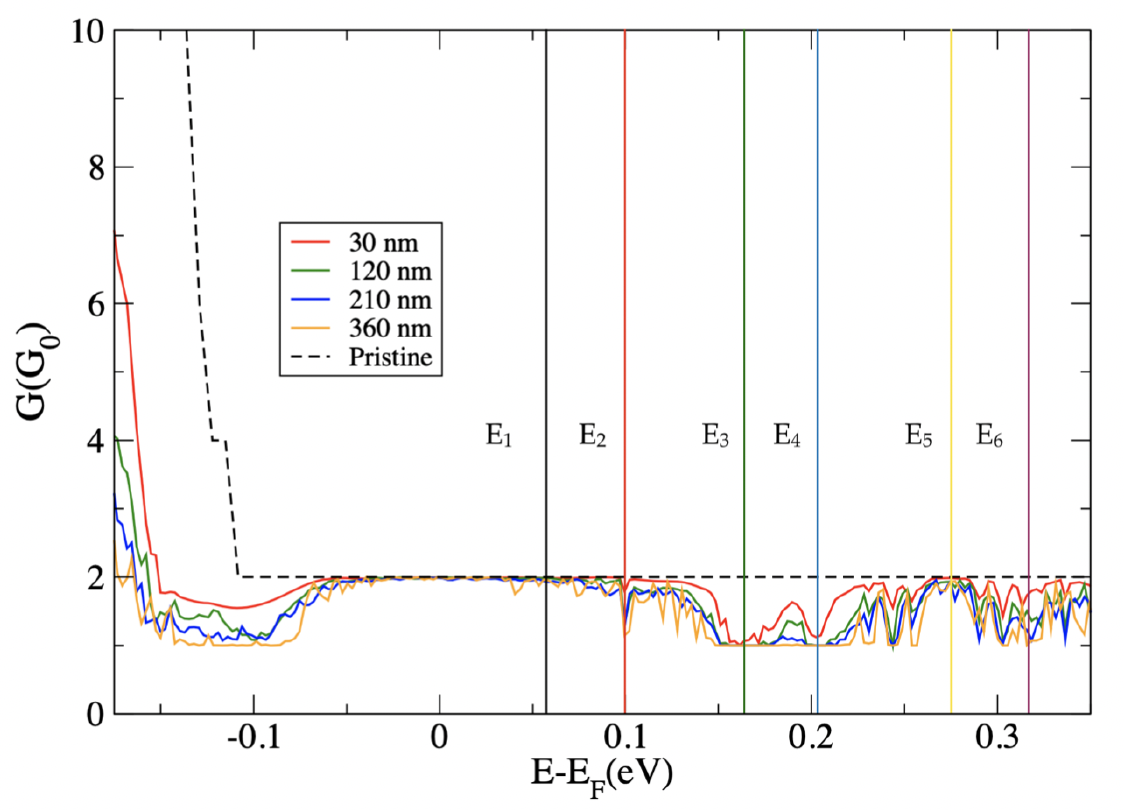}
    \caption{Conductance curves displaying the effect of having an increasing number of Fe adatoms.}
    \label{fig:drop_cond_fe}
\end{figure}

To gain a deep understanding on the spin transport, we consider the localization effect on the topological edge states in the presence of Iron atoms. Here considering random distribution of Fe atoms close to one edge, in this way we have that the conductance for the adatoms' free edge will be fixed as $1$, whereas backscattering events take place solely on impurity doped edge. Additionally, we have increased the scattering region keeping the same Fe linear concentration. In Fig.~\ref{fig:drop_cond_fe} we can see that this is the case, (i) the conductance never decreases below one, since only one edge is interacting with the magnetic adatoms, and (ii) conductance drop around $0.2$\,eV (the resonant Fe defect level, Fig.~\ref{fig:conductance_one_adatom}) became energetically spread with the increase of scattering region. Particularly, above $360$\,nm of scattering region the Fe doped edge transport become completely suppressed.

\begin{figure}
    \centering
    \includegraphics[width=\linewidth]{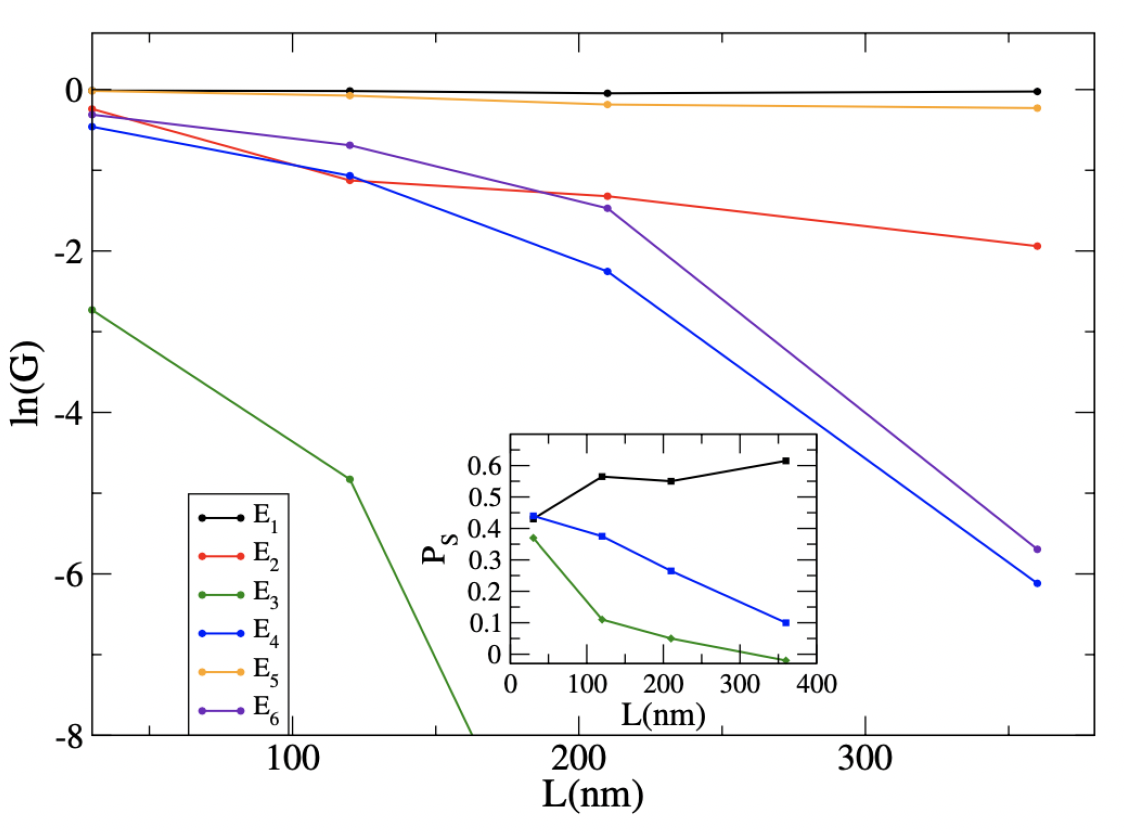}
    \caption{Logarithimic relation between the conductance and the length of the nanoribbon containing a distribution of impurities along one only edge. The inset represents the drop in the polarization for three energy points such that the we can relate this to the spin decoherence in our system. $\ln (G)$ vs sample's length (L) for different  energies chosen in the bulk gap. The inset shows the spin polarization ($P(L)$).}
    \label{log_condu}
\end{figure}

\begin{table}
\begin{ruledtabular}
\caption{\label{table:coherence_length} Table containing the localization lengths ($\xi$) for different energies following the linearity relation between $\ln (G)$ and the samples length (L).}
 \begin{tabular}{cc} 
 Energy label & $\xi$ (nm) \\ 
 \hline
 $E_1$  & 192 \\ 
 $E_3$  & 22 \\
 $E_4$  & 53 \\
 \end{tabular}
 \end{ruledtabular}
\end{table}

\begin{figure*}[!ht]
    \centering
    \includegraphics[width=0.9\linewidth]{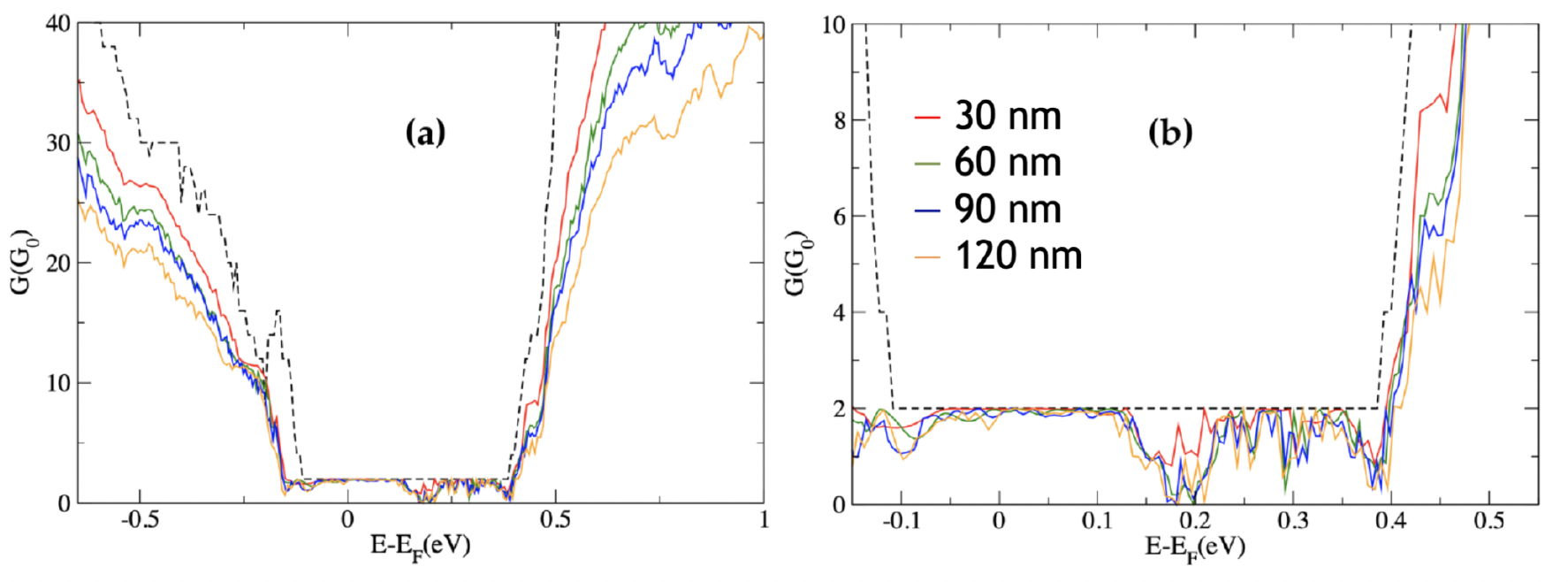}
    \caption{Conductance for Fe adatoms distributed along the edge of the nanoribbon(a) and (b) zoom-in view around the Fermi level. Each color represents different nanoribbon's lengths being proportional to the number of Iron impurities.}
    \label{fig:conductance_Fe}
\end{figure*}

This is similar to what is presented in Fig. \ref{fig:polarization_one_adatom} where only one of the polarization channels reaches negative values by virtue of TRS. One of our main results is related to the polarization as a function of the sample's length. Based on the same analysis performed in other two dimensional materials \cite{Grillo_2013, wudmir_spin_coherence, spin_coherence_pareek}, we can extract the values for the localization lengths obtained as a function of the conductance for a certain energy with respect to the device length. The localization length was obtained according to the following equation \cite{Datta1995}
\begin{equation}
\ln (G) = -\frac{L}{\xi},
\end{equation}
where $L$ is the sample's length, $G$ the conductance and $\xi$ the localization length. We show in Table \ref{table:coherence_length} the localization lengths obtained for three different energies within non-trivial topological gap. The localization length will quantify the penetration depth of the topological state within the scattering region. The behaviour of the conductance as a function of the scattering region length $L$ is depicted in Fig.~\ref{log_condu}. Here, we can see a long penetration depth (long localization length) for the energies non-resonant with the Fe impurity states. Particularly, we note that for the energy labeled as $E_4$ near 0.2 eV, the conductance almost vanishes, leading to a more localized nature ($\xi = 53$\,nm). Additionally, we see the inverse dependence of the polarization with the localization length [inset in Fig.~\ref{log_condu}]. That is, the spin transport is persistent in regions non-resonant with impurity states, even for the TRS breaking Fe adatoms.

\subsection{Bismuthene and multiple adatoms}

In the previous section we explored the effect of different adatoms adsorbed only close to one of the nanoribbons edge. Here, we can see the effect of distribution of the adatoms on the whole ribbon, keeping the same linear density of adatom. For Ni, given its preservation of TRS (non-magnetic) the backscattering protection is still preserved. This picture changes when Fe adatoms are attached to the nanoribbon as we can see in Fig. \ref{fig:conductance_Fe}. Besides the drop in the conductance for bulk states, we also can see how the breaking of TRS leads to a drop in the conductance within the topological energy gap. Additionally, we show the cumulative effect on increasing the length of the scattering region (keeping the same linear density of adatoms). That is, the larger the ribbon length, i.e. number of Fe adatoms in the structure, the larger is the number of possible backscattering events. It's worth mentioning that, since we are working in the diluted regime by considering just one impurity per block, most of the decrease in conductance appears to be confined to certain energy windows on which the impurity states are localized. We expect that, as the concentration increases, the drop in the conductance will start to become broader in energy. Despite the dilute regime, increasing the ribbon length the transport lose coherence \cite{Bass_2007, Radcliffe_1971, Das_sarma}.

\section{Conclusions}

In summary, we have shown that adatoms in topological materials can retain at some energy ranges the scattering forbidden character even for atoms breaking time-reversal symmetry. We characterize the resonance energy for magnetic impurities Fe and Co close to the topological insulator edge. After increasing the number sample's length, for impurities coupling only to one of the edges, we demonstrated that edge states began to localize as a result of the back-scattering from which we obtained the localization length. This certainly puts some bounds with respect to the dissipationless nature of the topological transport and would serve as a guiding rule in future measurements. Additionally we demonstrate the effect of having random distribution of magnetic adatoms on the topological insulator ribbon, where close to the Fermi energy the transport keep most of its character.

\begin{acknowledgments}
This work is partially supported by the Coordination for the Improvement of Higher Education Personnel - Brazil (CAPES) -
Finance Code 001, and São Paulo Research Foundation (FAPESP), grants no. 19/04527-0, 16/14011-2, 17/18139-6, and 17/02317-2. A. P. thanks to Prof. Alexandre R. Rocha for helpful discussions. The authors acknowledge the Brazilian National Scientific Computing Laboratory (LNCC), the Institute of Physics of the University of São Paulo (USP) and the Federal University of ABC (UFABC) for computational resources of the computers Santos Dumont, Josephson, and Titânio, respectively. 
\end{acknowledgments}


%

\end{document}